\documentclass[aps,twocolumn,pra,showpacs,nofootinbib]{revtex4-1}

\usepackage{graphicx}
\usepackage{dcolumn}
\usepackage{bm}
\usepackage{rotating}
\usepackage{enumerate}
\usepackage{longtable}
\usepackage{appendix}
\usepackage{multirow}
\usepackage{float}

\usepackage{ulem}
\renewcommand{\emph}[1]{\textit{#1}} 
\usepackage{graphics,epsfig,times,bbm,mathrsfs,color,hyperref}
\usepackage{amsmath}
\usepackage{amssymb}
\usepackage{amsfonts}
\usepackage{bbold}

\definecolor{orange}{rgb}{.7,0.5,0}
\newcommand{\be}{\begin{equation}}
\newcommand{\ee}{\end{equation}}
\newcommand{\Q}{QDD$_{N_1,N_2}\mbox{ }$}  
\newcommand{\lb}{\lambda}
\newcommand{\as}{\eta_x}
\newcommand{\bb}{\eta_y}
\newcommand{\cc}{\eta_z}
\newcommand{\tr}{\mbox{tr}}
\newcommand{\ket}[1]{|#1 \rangle} 
\newcommand{\bra}[1]{\langle #1 |} 
\newcommand{\DD}{D[\rho_{\text{S}}(T),\rho_{\text{S}}^0(T)]}
\newcommand{\is}{\ket{\psi}\bra{\psi}}
\newcommand{\vv}{\Vert}
\newcommand{\vm}{\vec{\mu}}
\newcommand{\vn}{\vec{\nu}}
\newcommand{\OO}{\mathcal{O}}
\newcommand{\K}{\mathcal{K}}

\newcommand{\bea} {\begin{eqnarray}}
\newcommand{\eea} {\end{eqnarray}}
\newcommand{\bes} {\begin{subequations}}
\newcommand{\ees} {\end{subequations}}

\newcommand{\green}[1]{\textcolor{black}{#1}}

\newcommand{\ma}[1]{\textcolor{black}{#1}}
\newcommand{\s}{\sigma}
\newcommand{\al}{\alpha}
\newcommand{\ignore}[1]{}

\begin{document}

\title{Rigorous Performance Bounds for Quadratic and Nested Dynamical Decoupling}

\author{Yuhou Xia}
\email{yxia@brynmawr.edu}
\affiliation{Departments of Mathematics and Physics, \\
Haverford College, Haverford,
Pennsylvania 19041, USA}

\author{G\"otz S. Uhrig}
\email{goetz.uhrig@tu-dortmund.de}
\affiliation{Lehrstuhl f\"{u}r Theoretische Physik I,
Technische  Universit\"{a}t Dortmund,
 Otto-Hahn Stra\ss{}e 4, 44221 Dortmund, Germany}

\author{Daniel A. Lidar}
\email{lidar@usc.edu}
\affiliation{
Departments of Electrical Engineering, Chemistry, and Physics\\
Center for Quantum Information Science \& Technology\\
University of Southern California, Los Angeles, California 90089, USA
}

\date{\rm\today}

\begin{abstract}
We present rigorous performance bounds for the quadratic dynamical decoupling 
(QDD) pulse sequence which protects a qubit from general decoherence, and for 
its nested generalization to an arbitrary number of qubits.  Our bounds apply 
under the assumption of instantaneous pulses
and of bounded perturbing environment and qubit-environment Hamiltonians such 
as those realized by baths
of nuclear spins in quantum dots.  We prove that if the total sequence time is 
fixed then the trace-norm distance between the unperturbed and protected system
 states can be made arbitrarily small by increasing the number of applied 
pulses. 
\end{abstract}



\maketitle

\section{Introduction}
The coupling between a quantum system and its environment typically causes 
decoherence, which is detrimental in quantum information processing (QIP) as it
 results in computational errors \cite{Nielsen:book}.  Over the years, 
various ways of suppressing quantum decoherence have been explored; see, e.g., 
Ref.~\cite{ByrdLidar:02a}.  The methodology we study here is dynamical 
decoupling (DD), which utilizes sequences of strong pulses to decouple the 
system from its environment 
\cite{Viola:98,Ban:98,Zanardi:98b,Viola:99,ByrdLidar:01}.  Recently, an optimal
 DD pulse sequence was discovered for the suppression of pure dephasing or 
longitudinal relaxation of a qubit coupled to a bath with a hard high 
frequency cutoff: Uhrig DD (UDD)  
\cite{Uhrig:07,Lee:08,Uhrig:08,Yang:08,Biercuk:09,biercuk-2009}. In UDD, the
instants $t_j$ ($j\in\{1,2\ldots N\}$) at which $N$ instantaneous $\pi$
pulses are applied are given by $t_j=T\lb_j$, where $T$ is the total time
of the sequence, and 
\begin{equation}  
\label{udd}
\lb_j =\sin^2\frac{j\pi}{2N+2}.
\end{equation}
By optimal it is meant that each additional pulse suppresses
dephasing or longitudinal relaxation
 to one additional order in an expansion in powers of $T$, i.e., $N$ pulses
reduce dephasing or longitudinal relaxation to $\mathcal{O}(T^{N+1})$. Rigorous 
performance bounds were established in Ref.~\cite{UL:10}. In this work we 
derive performance bounds for more general pulse sequences.

A near-optimal way to suppress general single-qubit decoherence, as opposed to 
only pure dephasing or pure longitudinal relaxation, is the quadratic DD (QDD) 
sequence \cite{WFL:09,NUDD,QL:11,KL:11,Jiang:11}.  A QDD sequence is obtained 
by nesting two UDD sequences of pulses which
are orthogonal in spin space. When these two UDD sequences comprise the 
same number of pulses, $N$, a QDD sequence of $(N+1)^{2}$
pulses will suppress general qubit
decoherence to $\mathcal{O}(T^{N+1})$, which is known from brute-force
symbolic algebra solutions for small $N$ to be near-optimal \cite{WFL:09}. It 
has been analytically proven that when the two UDD sequences comprise 
different numbers of pulses, $N_1$ and $N_2$, a QDD sequence of $(N_1+1)(N_2+1)$
pulses will suppress general qubit
decoherence to $\mathcal{O}(T^{\min(N_1,N_2)+1})$, and this is universal, i.e., 
holds for arbitrary baths and system-bath interactions 
\cite{NUDD,KL:11,Jiang:11}. Moreover, the dependence of the suppression order 
of each single qubit error type (dephasing, bit flip, or both) on $N_1$ and 
$N_2$ has also been established in detail both in terms of analytical bounds 
\cite{KL:11} and numerical simulations {\cite{QL:11,pasin11b}}. 

When the nesting process of UDD sequences is continued, one has the nested UDD 
(NUDD) sequence, which can protect multiple qubits, or general multi-level 
quantum systems, against general decoherence \cite{NUDD,Jiang:11}. It has been 
proven that NUDD is universal, and will suppress general multi-qubit
decoherence to $\mathcal{O}(T^{\min_i(N_i)+1})$, where $N_i$ are the orders of the 
UDD sequences being nested \cite{Jiang:11}. However, it is known that NUDD is 
sub-optimal \cite{NUDD}.

Our main results in this paper are analytical upper bounds, for QDD and NUDD, on
 the trace-norm distance between the states of the DD-protected qubit or 
qubits, and the unperturbed qubit or qubits, given as a function of the total 
evolution time and the norms of the bath operators.  Under the assumption that
 the bath operators have finite norms, the upper bound for NUDD shows that the 
trace-norm distance can be made arbitrarily small as a function of the minimal 
decoupling order $N$ of the UDD sequences comprising the NUDD sequence.
In the QDD case, a tighter bound is obtained by having different decoupling 
orders for different types of decoherence errors, using the results of 
Ref.~\cite{KL:11}.

The structure of this paper is as follows. We develop our results for QDD and 
NUDD in parallel, always starting with the simpler case of QDD. We first review
 the QDD and NUDD sequences in Section~\ref{sec:model}. We also derive bounding
 series for the two sequences in this section. In Section~\ref{sec:bounds} we 
use the bounding series in order to find explicit upper bounds on the different
 single-axis errors for QDD, and similar explicit upper bounds on different 
error types for NUDD. These results are then used in Section~\ref{sec:disbound}
 to derive the main results of this paper: trace-norm distance upper bounds for
 QDD and NUDD. We conclude in Section~\ref{sec:conclusions}, and present 
additional technical details in \ma{the Appendix}.

\section{Model}
\label{sec:model}

In this section we give a formal description of the QDD and NUDD sequences, and
 the decoherence they suppress.

\subsection{QDD}
\label{subsec:model-qdd}

We describe QDD as a nesting of two UDD sequences. The inner UDD sequence, 
denoted UDD$_{N_1}$, comprises $N_1$ $Z$-type pulses, meaning $N_1$ 
instantaneous rotations by $\pi$ about the $z$-axis of the qubit Bloch sphere, 
i.e., $Z=ie^{-i(\pi/2) \sigma_z}$. Similarly, the outer UDD sequence, denoted 
UDD$_{N_2}$, comprises $N_2$ $X$-type pulses, where $X=ie^{-i(\pi/2) \sigma_x}$.
  We denote the resulting QDD sequence by QDD$_{N_1,N_2}$.  To make sure that the
 qubit state is unaltered by the sequence itself, we append an additional 
pulse at the conclusion of the sequence if $N_1$ or $N_2$ is odd.  Thus, if 
$N'_{i}$, denotes the number of pulses in UDD$_{N_i}$, where $i \in \{1,2\}$, 
then
\be
\label{nopulses}
N'_{i} = N_i + N_i \mbox{ mod } 2 .
\ee
We define the dimensionless relative time $s:=t/T$, so that the $X$-type 
pulses are applied at times
\be
\label{xpulses}
\lb_j =\sin^2\frac{j\pi}{2N_2+2}
\ee
where $j=1,2,\ldots,N'_2$ and the $Z$-type pulses are applied at times
\be
\label{zpulses}
\lb_{j,k} = \lb_{j-1}+ (\lb_j-\lb_{j-1})\sin^2\frac{k\pi}{2N_1+2}
\ee
where $k = 1,2,\ldots,N'_1$.

We consider the most general form of ``bounded" time-independent single-qubit 
decoherence, which is described by the Hamiltonian
\begin{eqnarray}
\label{hamil-gendec1}
\widetilde H 
=\sum_{\alpha\in\{0,x,y,z\}} \sigma_\alpha \otimes B_\alpha
\end{eqnarray}
where {$\sigma_0=\mathbb{1}_{\text{S}}$}
 is the identity operator on the system.  
The operators $B_\alpha$ with $\alpha\in\{0,x,y,z\}$
are arbitrary except for the requirement that their sup-operator norms,
{i.e., the largest eigenvalue of $(B_\alpha^\dagger B_\alpha)^{1/2}$,}
 are  finite
\begin{equation}
\label{Jbounds}
J_\alpha:=\| B_\alpha \| < \infty.
\end{equation}

\begin{figure}[ht]
\begin{center}
\includegraphics[scale=.8,trim = 45mm  215mm 50mm 40mm]{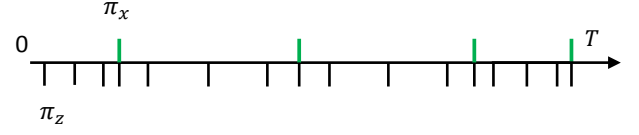} 
\end{center}
\caption{(Color online) Illustration of QDD$_{3,3}$. 
Time flows from left to right. The switching instants are 
distributed according to Eqs.~\eqref{xpulses} and \eqref{zpulses}. The 
pulses $\pi_z$ and $\pi_x$ are rotations about the $z$ and $x$ axes, 
respectively, by the angle $\pi$. The total sequence time is $T$.
}
\label{fig:qddseq}
\end{figure}

To decouple the qubit from the bath, we apply the \Q sequence, assuming 
Eq.~\eqref{hamil-gendec1}. See Fig.~\ref{fig:qddseq} for a schematic depiction
 of a QDD$_{3,3}$ sequence.

The \Q sequence is generated by the control Hamiltonian
\be
\label{conthaml}
H_c(s) = \frac{\pi}{2}{\sigma_x}
\sum_{j=1}^{N'_2}\delta(s-\lb_j)+
\frac{\pi}{2}{\sigma_z}
\sum_{j=1}^{N_2+1}\sum_{k=1}^{N'_1}\delta(s-\lb_{j,k}).
\ee
The corresponding control time-evolution operator is 
\be
\label{contevol}
U_c(s) = {\cal T} \exp\left( -i\int_0^s H_c(s')ds'\right),
\ee
where $\cal T$ is the time-ordering operator.  Thus, the toggling-frame 
Hamiltonian reads
\begin{subequations}
\label{togfrhaml}
\begin{eqnarray}
\label{conj}
H(s) & = & U^\dagger_c(s) {\widetilde H}U_c(s)\\
\label{hamil-qdd2}
& = & \sum_{\alpha \in \{0,x,y,z\}}f_\alpha(s) \sigma_\alpha \otimes B_\alpha ,
\end{eqnarray}
\end{subequations}
where the switching functions are 
\begin{subequations}
\label{modfunc}
\begin{eqnarray}
\label{0modfunc}
f_0{(s)} &=& 1  \hspace{26.5mm} {s\in}[0,1],\\
\label{xmodfunc}
f_x(s) &=& (-1)^{k-1} \hspace{15.8mm} {s\in}[\lb_{j,k-1},\lb_{j,k}),\\ 
\label{ymodfunc}
f_y(s) &=& (-1)^{k-1}(-1)^{j-1} \hspace{3.5mm} {s\in}[\lb_{j,k-1},\lb_{j,k}),\\
\label{zmodfunc} 
f_z(s) &=& (-1)^{j-1}  \hspace{16mm} {s\in}[\lb_{j-1},\lb_j).
\end{eqnarray}
\end{subequations}
For later use, we note that 
\be
f_y(s)=f_x(s)f_z(s).
\label{eq:fy}
\ee

Next, we consider the total time evolution given by the unitary operator
\begin{equation}
\label{Udef-qdd}
U(T) = {\cal T}\exp\big( -iT\int_0^1 H(s)ds\big).
\end{equation}
Standard time dependent perturbation theory provides
the following Dyson series for $U(T)$ 
\begin{subequations}
\label{expansion-relax}
\begin{eqnarray}
\label{Useries}
U(T) &=& \sum_{n=0}^\infty (-iT)^n 
\sum_{\{\vec{\alpha};\mathrm{dim}(\vec{\alpha})=n\}} 
f_{\vec{\alpha}}  \widehat Q_{\vec{\alpha}}
\\
\label{f_def}
f_{\vec{\alpha}} &:=& 
\int_0^1 ds_n f_{\alpha_n}(s_n) \int_0^{s_n} ds_{n-1} f_{\alpha_{n-1}}(s_{n-1})\ldots 
\nonumber\\ 
&&\int_0^{s_3} ds_2  f_{\alpha_2}(s_{2}) \int_0^{s_2} ds_1 f_{\alpha_1}(s_1)
\\
\label{q_def1}
\widehat 
Q_{\vec{\alpha}} &:=& (\sigma_{\alpha_n} \otimes B_{\alpha_n})   \cdots 
(\sigma_{\alpha_1} \otimes B_{\alpha_1})\\
& = & r\sigma_0^{n_0}\sigma_x^{n_x}\sigma_y^{n_y}\sigma_z^{n_z} 
\otimes B_{\alpha_n}
\cdots B_{\alpha_1}.
\label{q_def2}
\end{eqnarray}
\end{subequations}
The notation $\mathrm{dim}(\vec{\alpha})=n$ in Eq.~\eqref{Useries} means that 
the vector $\vec{\alpha}$ is restricted to having $n$ component 
$\{\alpha_n,\dots,\alpha_1\}$. The components are $\alpha_k \in \{0,x,y,z\}$ 
for $1\leq k \leq n$. Each integer $n_\alpha\ge 0, \alpha\in\{0,x,y,z\}$ counts 
how many times the operator $\sigma_\alpha \otimes B_\alpha$ appears in 
$\widehat Q_{\vec{\alpha}}$.  For given dimension $n$ of $\vec{\alpha}$, we have 
$n=n_0+n_x+n_y+n_z$. In this way, the complete sum over all possible sequences 
of $B_0$, $B_x$, $B_y$, and  $B_z$ are considered.  The rearrangement of the 
non-commuting product of Pauli matrices in Eq.~\eqref{q_def1} into the 
canonical form \eqref{q_def2} gives rise to 
{$r=(-1)^v$ where $v$ counts the number of times two different
Pauli-matrices have to pass each other based on 
$\sigma_\alpha\sigma_\beta = -\sigma_\beta \sigma_\alpha$ if $\alpha\neq\beta$.}

Our goal is to find an upper bound for each term 
$f_{\vec{\alpha}} \widehat Q_{\vec{\alpha}}$ separately.
We exploit $|f_\alpha(s)|= 1$, $\Vert\sigma_\alpha\Vert=1$, and
\bes
\bea
\Vert \widehat Q_{\vec{\alpha}} \Vert &\le& J_{\vec{\alpha}}
\\
J_{\vec{\alpha}} &:=& J_{\alpha_n} J_{\alpha_{n-1}} \ldots J_{\alpha_1}
\eea
\ees
to obtain the upper bounding series given  below 
for the series \eqref{Useries} term  by term
\bes
\label{eq:ubound}
\bea
\label{ubound_1}
\| U(T) \| &\leq & \sum_{n=0}^\infty T^n F_n
\sum_{\{\vec{\alpha};\mathrm{dim}(\vec{\alpha})=n\}}  
J_{\vec{\alpha}} \\
\label{S_def}
&=:& S_\text{Q}(T)\\
\label{ubound_3}
&=& \exp[(J_0+J_x+J_y+J_z)T],
\eea
\ees
where
\be
F_n := \int_0^1 ds_n  \int_0^{s_n} ds_{n-1}\ldots 
\int_0^{s_3} ds_2   \int_0^{s_2} ds_1 =\frac{1}{n!}.
\label{eq:Fn}
\ee
The equality between \eqref{ubound_1} and \eqref{ubound_3} is
most easily seen by realizing that the right hand side of \eqref{ubound_1}
is the Dyson series of \eqref{ubound_3} as obtained by
time dependent perturbation theory.
This series will be used extensively when we compute the performance bounds in 
Section~\ref{sec:bounds}  which generalize the UDD results in 
Ref.\ \cite{UL:10}.

\subsection{NUDD}
\label{subsec:model-nudd}

The NUDD sequence is a generalization of the UDD and QDD sequences that 
suppresses decoherence for a multi-qubit system.  Assume that the system 
comprises $m$ qubits. We consider the most general form of ``bounded" 
time-independent $m$-qubit decoherence. Subject only to the  constraint on the 
bath operators $B_{\vm}$ that they are bounded
\be
\label{nbound}
J_{\vm}:=\Vert B_{\vm}\Vert {< \infty},
\ee
the Hamiltonian that gives rise to this general decoherence is 
\be
\label{nuddmodel}
\widetilde{H} =\sum_{\vec{\mu}\in{D_m}}\widehat{\sigma}_{\vm}\otimes B_{\vm}
\ee
where 
\bes
\label{sigma_D}
\bea
{D_m} &{:=}& {\{(0,0),(1,0),(1,1),(0,1)\}^m}
\\
\widehat{\sigma}_{\vm} &:=& \sigma_{1,\mu_1}\otimes\cdots\otimes\sigma_{m,\mu_m}.
\label{widesig}
\eea
\ees
Here the $\sigma_{j,\mu_j}$'s are Pauli matrices or the identity and we use 
$j\in\{1,\dots,m\}$ to index the qubits.  For the $j$th qubit, we use the 
binary component ${\mu_j}\in \{(0,0),(1,0),(1,1),(0,1)\}$ 
of the vector $\vec{\mu}:=(\mu_1,\mu_2,\ldots,\mu_m)$ to denote 
the Pauli matrix subscripts $\{0,x,y,z\}$, respectively.

The NUDD sequence for an $m$-qubit system consists of $2m$ nested levels {(two per qubit)}.  
Let $T$ be the total duration of the NUDD sequence and $N'_i$ be the 
decoupling order of the $i$th-level UDD [Eq.~\eqref{nopulses}]. Then the 
pulses at the $i$th level are applied at the instants
\begin{eqnarray}
\nonumber
\lefteqn{\lambda_{l_{2m},l_{2m-1},...,l_i}}\\
\nonumber
& = &\lambda_{l_{2m},l_{2m-1},...,l_{i+1}}+(\lambda_{l_{2m},l_{2m-1},...,l_{i+1}+1}\\
\label{time-indices}
& &{}-\lambda_{l_{2m},l_{2m-1},...,l_{i+1}})\sin^2\frac{l_i\pi}{2 N'_i+2}
\end{eqnarray}
where $1\leq i \leq 2m$ and $1\leq l_i \leq N'_i$.  {Even values of $i$ correspond to $\sigma_x$ pulses applied to qubit number $j=i/2$, while odd values of $i$ correspond to $\sigma_z$ pulses applied to qubit number $j=(i+1)/2$.} The control Hamiltonian is 
therefore
\be
\begin{split}
& H_c(s) = \frac{\pi}{2}\sum^{{2m}}_{\{i{\geq 2}\textrm{, even}\}}{{\sigma_{i/2,x}}}
\sum_{\{{l_{{p}}=1\}_{{{p}}=i}^{{2m}}}}^{N'_{{p}}}\delta(s-\lambda_{l_{{2m}},...,l_i}) \\
\label{ncon-haml}
&\quad +\frac{\pi}{2} \sum^{{2m-1}}_{\{i{\geq 1}\textrm{, odd}\}}{{\sigma_{(i+1)/2,z}}}
\sum_{\{{l_{{p}}=1\}_{{{p}}=i}^{2m}}}^{N'_{{p}}} \delta(s-\lambda_{l_{{2m}},...,l_i}) ,
\end{split}
\ee
{where the inner sums are multiple sums, one for each value of the index $p$} {labeling the nesting levels $i, i+1, ..., 2m$.} The control time-evolution operator is 
\be
\label{ncontevol}
U_c(s) = {\cal T} \exp\big( -i\int_0^s H_c(s')ds'\big).  
\ee
The toggling-frame Hamiltonian is
\begin{subequations}
\begin{eqnarray}
H(s)&=&U_c(s)^\dagger \widetilde{H}U_c(s) \\
&=&\sum_{\vm\in{D_m}} f_{\vm}{(s)}\mbox{ }
\widehat{\sigma}_{\vm} \otimes B_{\vm} \\
\label{fmu}
f_{\vm}{(s)} &=&\prod_{j=1}^m f_{j,\mu_j} {(s)}
\end{eqnarray}
\end{subequations}
where the $f_{j,\mu_j}(s)$'s are the switching functions.  
They can be obtained by the anticommutation relations of the Pauli matrices
\bes
\bea
f_{j,(0,0)}{(s)} &=& 1
\\ \nonumber 
&& \hspace*{-5mm} {\text{for}\  s\in [0,1] }
\\
f_{j,(1,0)}{(s)} &=& {(-1)^{l_{2j-1}}} 
\\ \nonumber 
&& \hspace*{-5mm} {\text{for}\ s\in 
[\lambda_{l_{2m},l_{2m-1},...,l_{2j-1}-1},\lambda_{l_{2m},l_{2m-1},...,l_{2j-1}})}
\\
f_{j,(0,1)}{(s)} &=&
{(-1)^{l_{2j}}} 
\\ \nonumber 
&& \hspace*{-5mm} {\text{for}\ s\in 
[\lambda_{l_{2m},l_{2m-1},...,l_{2j}-1},\lambda_{l_{2m},l_{2m-1},...,l_{2j}})}
\\
f_{j,(1,1)}{(s)}&=& f_{j,(1,0)}{(s)}f_{j,(0,1)}{(s)} 
\\ \nonumber 
&&  \hspace*{-5mm} {\text{for}\  s\in [0,1]}.
\eea
\ees
In the last equation we used Eq.~\eqref{eq:fy}.

Now consider the total time evolution given by the unitary operator
\begin{equation}
\label{Udef-nudd}
U(T) = {\cal T}\exp\big( -iT\int_0^1 H(s)ds\big).
\end{equation}
Standard time dependent perturbation theory provides
the following Dyson series for $U(T)$ 
\begin{subequations}
\label{nexpansion-relax}
\begin{eqnarray}
U(T) &=& \sum_{n=0}^\infty (-iT)^n 
\sum_{\{\vm_k{\in D_m}\}_{k=1}^n} 
f_{\vm_1,...,\vm_n} \; \widehat Q_{\vm_1,...,\vm_n}\notag \\
\label{nUseries}\\
f_{\vm_1,...,\vm_n} &:=& 
\int_0^1 ds_n f_{\vm_n}(s_n) \int_0^{s_n} ds_{n-1} f_{\vm_{n-1}}(s_{n-1})\ldots 
\nonumber\\ 
&&\int_0^{s_3} ds_2  f_{\vm_2}(s_{2}) \int_0^{s_2} ds_1 f_{\vm_1}(s_1)
\label{nf_def}
\\
\widehat 
Q_{\vm_1,...,\vm_n} &:=& (\widehat{\sigma}_{\vm_n} \otimes B_{\vm_n})   \cdots 
(\widehat{\sigma}_{\vm_1} \otimes B_{\vm_1})
\label{nq-def1}
 \\
& = & r \prod_{\vm_k\in D_m} \widehat{\sigma}_{\vm}^{n_{\vm}} \otimes 
B_{\vm_n}
\ma{\cdots}B_{\vm_1},
\label{nq_def2}
\end{eqnarray}
\end{subequations}
where $r=\prod_{j=1}^m r_j$ and each 
{$r_j:=(-1)^{v_j}$,} 
{where $v_j$ counts how often two different $\sigma_{j,\mu_j}$ 
{with the same $j$} pass each other 
in order to obtain \eqref{nq_def2} from \eqref{nq-def1}}. 
 Each $n_{\vm}$ in Eq.~\eqref{nq_def2} indicates how many times 
$\widehat{\sigma}_{\vm}\otimes B_{\vm}$ appears in Eq.~\eqref{nq-def1} and the 
$n_{\vm}$'s satisfy $\sum_{\vm\in{D_m}} n_{\vm}=n$.

Following steps  analogous to the QDD case we have
\bes
\bea
\label{nf-bd}
{|f_{\vm}|}&{=}&{1}
\\
\label{nq-bd}
\Vert \widehat Q_{\vm_1,...,\vm_n} \Vert &\leq& \prod_{\vm\in{D_m}} J_{\vm}^{n_{\vm}}.
\eea
\ees
{This allows us to write}
\bes
\bea
\label{UN-bound}
\Vert U(T) \Vert &\leq &  \sum_{n=0}^\infty T^n F_n
\sum_{\{\vm_k{\in D_m}\}_{k=1}^n} 
\prod_{\vm\in{D_m}} J_{\vm}^{n_{\vm}}
\\
&=:& S_{\rm N}(T),
\eea
\ees
where $F_n$ was defined in Eq.~\eqref{eq:Fn} and the right hand side of 
\eqref{UN-bound} is the Dyson series applied to
\be
\label{nbseries}
S_{\rm N}(T) =\exp\Big(T\sum_{\vm\in {D_m}}J_{\vm}\Big).
\ee
For an alternative derivation see Section
\ref{subsec:bound-nudd-operator}.

\section{Bounds for General Decoherence}
\label{sec:bounds}

\subsection{QDD}
\label{subsec:bound-qdd-operator}

Very recently Ref.~\cite{KL:11} found rigorous lower bounds on the decoupling 
orders of a QDD$_{N_1,N_2}$ sequence.  The result is summarized in 
Table~\ref{qdd-decpl}.  We define the decoupling order of a single-axis error 
$\sigma_\alpha$ to be $d_\alpha$ for $\alpha \in \{x,y,z\}$.  Thus, after 
applying the QDD$_{N_1,N_2}$ sequence, the first non-vanishing term in the 
Dyson series~\eqref{Useries} that contributes to the $\sigma_\alpha$-type error 
is of order $T^{d_\alpha+1}$ or higher.\footnote{In Ref.~\cite{QL:11} the result 
for the decoupling order of $\sigma_z$ when $N_1$ is odd was found numerically 
(using a spin-bath model) to be $\min\{2N_1+1,N_2\}$, which improves on the 
analytical bound $\min\{N_1+1,N_2\}$ from Ref.~\cite{KL:11}. 
In all other cases  Refs.~\cite{QL:11,KL:11} were in perfect agreement.} 
We will use this result in deriving the bounds for decoherence.  
\begin{table}[ht]
\begin{tabular}{|c|c|c|c|}\hline
Single-axis error {$\sigma_\alpha$}
& $N_1$ mod 2 &$N_2$ mod 2 & Decoupling order {$d_\alpha$}
\\\hline
{$\alpha=x$} & 0 or 1& 0 or 1& $N_1$
\\\hline
\multirow{4}{*}{{$\alpha=y$}}& 0&0& max\{$N_1$,$N_2$\}
\\
&0&1&max\{$N_1+1$,$N_2$\}
\\
&1&0&$N_1$\\
&1&1&$N_1+1$\\\hline
\multirow{2}{*}{{$\alpha=z$}}&0&0 or 1& $N_2$\\
&1&0 or 1&min\{$N_1+1,N_2$\}\\\hline
\end{tabular}
\caption{Summary of single-axis error suppression from Ref.~\cite{KL:11}.}
\label{qdd-decpl}
\end{table}

Any operator acting on the qubit subspace 
can be expanded in terms of the Pauli matrices and the identity. Hence
\bea
U(T)=\sum_{\alpha\in\{0,x,y,z\}} \sigma_\alpha\otimes A_\alpha(T)
\label{Usplit}
\eea
is another way to write Eq.~\eqref{Useries}, which serves to define the bath 
operators $A_\alpha(T)$.
We classify the decoherence error based on the parities
\be
p_\alpha:=n_\alpha \textrm{ mod } 2,\quad  
\alpha\in\{x,y,z\}.
\label{eq:p}
\ee
Using 
\be
\sigma_\alpha \sigma_\beta = 
i\varepsilon_{\alpha \beta\gamma}\sigma_\gamma + \delta_{\alpha \beta} 
{\mathbb{1}}
\quad \alpha,\beta\in\{x,y,z\}
\label{eq:sasb}
\ee
where $\varepsilon_{\alpha \beta\gamma}$ is the Levi-Civita symbol,
we find the results gathered in Table \ref{tab:parity}.

\begin{table}[ht]
\begin{tabular}{|c|c|c|c|c|c|}
\hline
case $j$ & $p_x$ & $p_y$ & $p_z$ & Channel & Decoupling order\\
\hline
{0} & {0} & {0} & {0} & {$\sigma_0 $} & {$1$}\\
\hline
1 & 0 & 0 & 1& $ \sigma_z$ & $d_z$\\
\hline
2 & 0 & 1 & 0& $ \sigma_y$ & $d_y$\\
\hline
3 & 0 & 1 & 1& $ \sigma_x$ & $d_x$\\
\hline
4 & 1 & 0 & 0& $ \sigma_x$ & $d_x$\\
\hline
5 & 1 & 0 & 1& $ \sigma_y$ & $d_y$\\
\hline
6 & 1 & 1 & 0& $ \sigma_z$ & $d_z$\\
\hline
{7} & {1} & {1} & {1} & {$\sigma_0$} & {$1$}\\
\hline
\end{tabular}
\caption{Classification of the error terms in the series Eq.~\eqref{Useries}
according to the parities $p_\alpha:=n_\alpha \text{ mod } 2$. 
Recall that $n_\alpha$ 
counts how many times the operator $\sigma_\alpha \otimes B_\alpha$ appears in 
$\widehat Q_{\vec{\alpha}}$. The Channel column gives the corresponding 
single-axis error term in Eq.~\eqref{Usplit}. The last column gives the order 
to which the single-axis error is suppressed. 
For completeness, the identity channel $\sigma_0$ is also 
listed, though rather than being suppressed
it increases in linear order.
\label{tab:parity}}
\end{table}

Similarly to the bounding series found in Ref.~\cite{UL:10}
we can straightforwardly deduce bounding series for the various cases
in Table~\ref{tab:parity} from $S_{\rm Q}(T)$ in Eq.~\eqref{S_def}. We first 
write $S_{\rm Q}(T)$ as a sum over 
the eight terms differing in at least one parity
\be
S_{\rm Q}(T) = \sum_{j=0}^7 S_j(T), \quad j = 2^0 p_z +2^1 p_y + 2^2 p_x.
\label{S_Q}
\ee
The parity $p_\alpha$ of $n_\alpha$ is the parity of $S_{\rm Q}(T)
=\exp[(J_0+J_x+J_y+J_z)T]$  as function of $J_\alpha$.
Since $\exp(J_\alpha T)=\cosh(J_\alpha T) + \sinh(J_\alpha T)$
nicely splits even and odd contributions, we can split $S_\text{Q}(T)$ 
into separate bounding functions as listed in Table~\ref{tab:bounds}.

\begin{table}[hb]
\begin{tabular}{|c|c|c|c|c|}
\hline
case $j$ 
& $p_x$ & $p_y$ & $p_z$ & Bounding function $S_j$\\
\hline
{0} & {0} & {0} & {0} & 
{$ e^{J_0T} \cosh(J_xT)\cosh(J_yT)\cosh(J_zT)$} \\
\hline
1 & 0 & 0 & 1& $ e^{J_0T} \cosh(J_xT)\cosh(J_yT)\sinh(J_zT)$ \\
\hline
2 & 0 & 1 & 0& $ e^{J_0T} \cosh(J_xT)\sinh(J_yT)\cosh(J_zT)$ \\
\hline
3 & 0 & 1 & 1& $ e^{J_0T} \cosh(J_xT)\sinh(J_yT)\sinh(J_zT)$ \\
\hline
4 & 1 & 0 & 0& $ e^{J_0T} \sinh(J_xT)\cosh(J_yT)\cosh(J_zT)$ \\
\hline
5 & 1 & 0 & 1& $ e^{J_0T} \sinh(J_xT)\cosh(J_yT)\sinh(J_zT)$ \\
\hline
6 & 1 & 1 & 0& $ e^{J_0T} \sinh(J_xT)\sinh(J_yT)\cosh(J_zT)$ \\
\hline
{7} & {1} & {1} & {1} & 
{$e^{J_0T} \sinh(J_xT)\sinh(J_yT)\sinh(J_zT)$} \\
\hline
\end{tabular}
\caption{Functions of the respective bounding series
 for the various contributions of parity combinations. 
}
\label{tab:bounds}
\end{table}

We would like to use the bounding functions $S_j$ to upper-bound $\|A_\al\|$. 
To this end, note first that, using Eqs.~\eqref{Usplit} and \eqref{eq:sasb}, 
for $\alpha\in\{x,y,z\}$
\bea
\frac{1}{2}\tr_{{\text{S}}}[\s_\al U(T)] = 
\frac{1}{2}\sum_\beta \tr[\s_\al\s_\beta]A_\beta (T) = A_\al (T),
\label{eq:tr_S}
\eea
where $\tr_{{\text{S}}}$ denotes the partial trace over the system. 
On the other hand, using Eq.~\eqref{expansion-relax},
\bea
\tr_{{\text{S}}}[\s_\al U(T)] = \sum_{n=0}^\infty (-iT)^n \!\!\!\!\!
\sum_{\{\vec{\alpha};\mathrm{dim}(\vec{\alpha})=n\}}\!\!\!\!\! 
f_{\vec{\al}}\tr_{{\text{S}}}[\s_\al \hat{Q}_{\vec{\al}}],
\eea
and
\bea
\tr_{{\text{S}}}[\s_\al \hat{Q}_{\vec{\al}}]=
r \tr[\s_\al \s_x^{n_x} \s_y^{n_y} \s_z^{n_z}]  
B_{\alpha_n}\cdots B_{\alpha_1}.
\eea
Using the fact that only the parity \eqref{eq:p} of the exponents matters, 
we can rewrite the trace as
\bes
\label{eq:theta}
\bea
\theta_{\al,\vec{p}} &:=& 
\frac{1}{2}{\Big|}\tr[\s_\al \s_x^{n_x} \s_y^{n_y} \s_z^{n_z}] 
{\Big|}
\\
&=& \frac{1}{2}{\Big|}\tr[\s_x^{p_x\oplus\delta_{\al,x}} 
\s_y^{p_y\oplus\delta_{\al,y}} \s_z^{p_z\oplus\delta_{\al,z}}]{\Big|},
\eea
\ees
where $\vec{p}:=(p_x,p_y,p_z)$ and $\oplus$ denotes addition modulo $2$. 
{The values of $\theta_{\al,\vec{p}}$ are given Table~\ref{tab:theta}.}
\begin{table}[ht]
\begin{tabular}{|c|c|c|c|c|}
\hline
$p_x\oplus\delta_{\al,x}$ & $p_y\oplus\delta_{\al,y}$ & 
$p_z\oplus\delta_{\al,z}$ & $\theta_{\al,\vec{p}}$ \\
\hline
0 & 0 & 0 & $1 $ \\
\hline
0 & 0 & 1& $ 0$ \\
\hline
0 & 1 & 0& $ 0$ \\
\hline
0 & 1 & 1& $ 0$ \\
\hline
1 & 0 & 0& $ 0$ \\
\hline
1 & 0 & 1& $ 0$ \\
\hline
1 & 1 & 0& $ 0$ \\
\hline
1 & 1 & 1 & 1 \\
\hline
\end{tabular}
\caption{The eight cases implied by Eq.~\eqref{eq:theta}.}
\label{tab:theta}
\end{table}

Combining Eqs.~\eqref{eq:tr_S}-\eqref{eq:theta}, we have, 
\be
\|A_\al (T)\| \leq  \sum_{n=0}^\infty T^n F_n \!\!\!\!\!
\sum_{\{\vec{\alpha};\mathrm{dim}(\vec{\alpha})=n\}} \!\!\!\!\!  
{\theta_{\al,\vec{p}}} J_{\vec{\al}} 
\label{eq:A_al-bound}
\ee
where $F_n$ was defined in Eq.~\eqref{eq:Fn}.

{In view of Table~\ref{tab:theta}, Eq.\ \eqref{eq:A_al-bound}
provides the decomposition  by parity of Eq.~\eqref{eq:ubound}.}
Consider, e.g., the case $\al=x$. Then Table~\ref{tab:theta} tells us that 
$\theta_{x,\vec{p}}=1$ only when $\{p_x=1,p_y=0,p_z=0\}$ (second row) or 
$\{p_x=0,p_y=1,p_z=1\}$ (bottom row). In all other cases $\theta_{x,\vec{p}}=0$. 
Comparing with Table~\ref{tab:bounds}, we see that these two non-zero cases 
correspond to $j=4$ and $j=3$, respectively. A similar argument informs us 
that $\theta_{y,\vec{p}}=1$ only when $\{p_x=0,p_y=1,p_z=0\}$ or 
$\{p_x=1,p_y=0,p_z=1\}$, which corresponds to $j=2$ and $j=5$ in 
Table~\ref{tab:bounds}, and $\theta_{z,\vec{p}}=1$ only when 
$\{p_x=0,p_y=0,p_z=1\}$ or $\{p_x=1,p_y=1,p_z=0\}$, which corresponds to $j=1$ 
and $j=6$ in Table~\ref{tab:bounds}.

Now, note that the right-hand side of Eq.~\eqref{eq:A_al-bound} without 
the $\theta_\al$ factor is just $S_{\rm Q}(T)$ as defined in Eq.~\eqref{S_def}. 
We can thus conclude from Eq.~\eqref{eq:A_al-bound} that, due to the 
$\theta_\al$ prefactor, its right-hand-side consists of only two non-vanishing 
terms for each value of $\theta_\al$, which acts like a Kronecker delta 
function for the parity triple $(p_x,p_y,p_z)$ of the bounding functions $S_j$, 
namely $\al=x \Rightarrow j=3,4$, $\al=y \Rightarrow j=2,5$, $\al=z 
\Rightarrow j=1,6$. Thus,
\bes
\label{eq:A_al-bounds}
\bea
\|A_x(T)\| &\leq& S_3(T)+S_4(T)   \\
\|A_y(T)\| &\leq& S_2(T)+S_5(T)    \\
\|A_z(T)\| &\leq& S_1(T)+S_6(T).
\eea
\ees

Next, to account for the suppression of decoherence by QDD we define 
\be
\label{taylor}
p_{k}^{(j)}:=\frac{1}{k!}\frac{\partial^k}{\partial T^k}S_j{\Big\vert}_{T=0} ,
\ee
so that
\be
\label{bseries-taylor}
S_j = \sum_{k=0}^\infty p_k^{(j)} T^k.
\ee
We consider the partial Taylor series 
$\Delta^{(j)}_d$ of the  analytic bounding functions $S_j$ which leave out 
the contributions up to and including $T^d$
\be
\label{eq:Delta-def}
\Delta_{{d}}^{(j)} := \sum_{k={d}+1}^\infty p_k^{(j)} T^k.
\ee
According to Tables~\ref{tab:parity} and \ref{tab:bounds}, the contributions
of case $j$ are bounded by the corresponding
partial Taylor series $\Delta_d^{(j)}$ where $d$ is the decoupling order of the 
channel. This is a manifestation of the fact that the QDD sequence causes the first 
$d_\alpha$ powers in $T$ of $A_\alpha$ to vanish, i.e., 
$A_\alpha(T)=\mathcal{O}(T^{d_\alpha+1})$  \cite{KL:11}.
Using Eq.~\eqref{eq:A_al-bounds} we deduce that
\begin{subequations}
\label{eq:Aperp}
\begin{eqnarray}
\label{eq:Ax}
L_x:= \Delta_{d_x}^{(3)} + \Delta_{d_x}^{(4)} &\geq& \Vert A_x(T)\Vert
 \\
\label{eq:Ay}
L_y:=\Delta_{d_y}^{(2)} + \Delta_{d_y}^{(5)}&\geq& \Vert A_y(T)\Vert   
\\
\label{eq:Az}
L_z:=\Delta_{d_z}^{(1)} + \Delta_{d_z}^{(6)}&\geq&\Vert A_z(T)\Vert.
\end{eqnarray}
\end{subequations}
This is our first key result for QDD.

Due to the analyticity in the variable $T$ of $S_j$ for each $j$, we know that 
the residual term vanishes for ${d} \rightarrow \infty$, that is 
\be
\label{vani-res}
\lim_{{d}\rightarrow \infty} \Delta_{{d}}^{(j)} = 0.
\ee

We define the dimensionless parameters
($\hbar$ is set to unity)
\begin{equation}
\varepsilon :=J_0T \quad
\eta_\alpha := J_\alpha/J_0
\end{equation}
where $\alpha\in\{ x,y,z\}$.  Thus, we can write the bounding functions $S_j$ 
as $S_j(\varepsilon,\vec{\eta})$, where $\vec{\eta}=(\eta_x,\eta_y,\eta_z)$.  
By Taylor-expanding $S_j(\varepsilon,\eta_x,\eta_y,\eta_z)$ we can express the 
bounding functions in terms of $\varepsilon$ and functions 
$\{g_l^{(n)}(\vec{\eta})\}_{n=1}^6$ given in Appendix~\ref{app:g}:
\begin{subequations}
\begin{eqnarray}
\label{delta-poly1}
\Delta_{d_\alpha}^{(j)}& = & \sum_{n=d_\alpha+1}^\infty 
g_n^{(j)}(\vec{\eta})\varepsilon^n\\
\label{delta-poly2}
& = & g_{d_\alpha+1}^{(j)}(\vec{\eta})\varepsilon^{d_\alpha+1}+
\OO(\varepsilon^{d_\alpha+2}).
\end{eqnarray}
\end{subequations}

\begin{figure}[H]
\begin{center}
\includegraphics[scale=.5,width=0.99\columnwidth]
{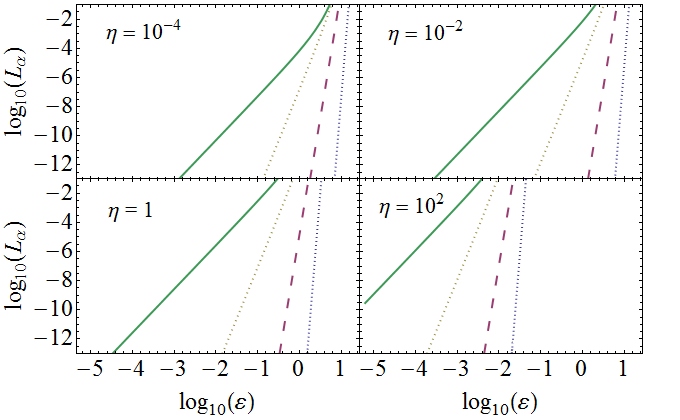} 
\end{center}
\caption{(Color online) The upper bound $L_\alpha$ for 
$\Vert A_x\Vert=\Vert A_y\Vert = \Vert A_z\Vert$ according to 
Eqs.~\eqref{eq:Aperp} in the isotropic case, i.e., $\eta_x=\eta_y=\eta_z$, 
is given as a function of $\varepsilon$ for various number of pulses 
$N_1=N_2=N\in \{2,6,16,34\}$, at fixed values of $\eta_\alpha$.  In each panel 
the curves become steeper as $N$ increases.}
\label{fig:xy1}
\end{figure}

The behaviors of the bounding functions $L_x,L_y$ and 
$L_z$ are shown in Fig.~\ref{fig:xy1} for different $\vec{\eta}$.  
Here we assume that the decoherence is isotropic, i.e., 
$\eta:=\eta_x=\eta_y=\eta_z$.  Given this 
assumption, it makes sense to choose $N_1=N_2$ so that the decoupling orders 
for different types of errors are close.  To simplify the computation, we 
choose both $N_1$ and $N_2$ to be even and thus, according to 
Table~\ref{qdd-decpl}, $d_x=d_y=d_z$.  Note that in this situation $L_x$, $L_y$ 
and $L_z$ are identical.

\begin{figure}[ht]
\begin{center}
\includegraphics[scale=.5,width=0.99\columnwidth]
{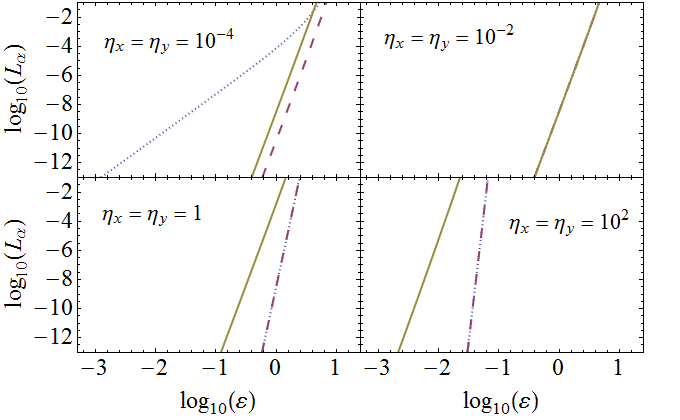} 
\end{center}
\caption{(Color online) The upper bounds for $\Vert A_x\Vert$ 
(dotted lines), $\Vert A_y\Vert$ (dashed lines) and 
$\Vert A_z\Vert$ (filled lines)  according to Eq.~\eqref{eq:Aperp} are 
depicted.  Here $N_2$ is $10$. 
The values of $N_1$ in the four panels going 
from left to right and top to bottom are $2,10,18$ and $34$.  
The value of $\eta_z=10^{-2}$.}
\label{fig:xy2}
\end{figure}

\begin{figure}[ht]
\begin{center}
\includegraphics[scale=.5,width=0.99\columnwidth]
{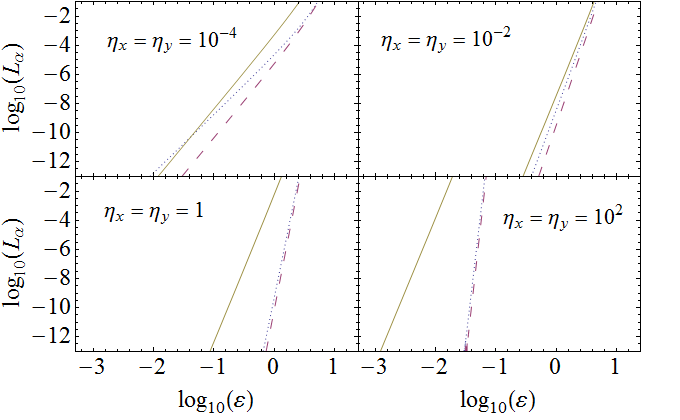} 
\end{center}
\caption{(Color online) The upper bounds for $\Vert A_x\Vert$ 
(dotted lines), $\Vert A_y\Vert$ (dashed lines) and 
$\Vert A_z\Vert$ (filled lines)  according to Eqs.~\eqref{eq:Aperp} are 
depicted.  We make $N_2$ to be $9$ in this case. 
The values of $N_1$ in the four panels going 
from left to right and top to bottom are $3,10,19$ and 
$34$. The value of $\eta_z=10^{-2}$ as in the previous case.}
\label{fig:xy4}
\end{figure}

We also investigated the case where $\eta_x=\eta_y \neq\eta_z$ and how the 
parities of $N_1$ and $N_2$ affect the bounds.  In Fig.~\ref{fig:xy2} 
and Fig.~\ref{fig:xy4}, we plotted two cases.  
The first is where both $N_1$ and $N_2$ are even, which are
 represented by the thick lines. The second is where $N_1$ and $N_2$ are either
 even or odd, which are represented by the thin lines.  We picked $\eta_z$ to 
be $10^{-2}$ for all the plots and $\eta_x=\eta_y\in\{10^{-4},10^{-2},1,10^2\}$.  
Since $\eta_z$ is fixed, we also fixed the values of $N_2$ for the two cases.  
We picked $N_2$ to be 10 in the first case and 9 in the second case to explore 
the effect of parity.  We varied the value of $N_1$ based on the values 
$\eta_x$ and $\eta_y$.  We can see from the figures that in the majority of 
cases, the parities of $N_1$ and $N_2$ only change the decoupling order by 1 
or do not change it at all, so the bounds do not vary much.  However, when 
$N_1$ is small compared to $N_2$, the parities can change the bounds 
nontrivially as seen in the top left panels of Fig.~\ref{fig:xy2} 
and Fig.~\ref{fig:xy4}.  We can also 
see this from Table~\ref{qdd-decpl}.  When $N_1$ is small compared to $N_2$, 
the decoupling orders of $\sigma_y$ and $\sigma_z$ can decrease considerably 
when $N_1$ switches between even and odd values.

Equation~\eqref{eq:Aperp}  allows us to establish tight
bounds on the representation of the unitary time evolution
as given in Eq.~\eqref{Usplit}.  In subsection \ref{subsec:D-bound} we use 
this fact to establish a bound on the trace-norm distance between the 
unperturbed qubit state and the protected qubit state.

\subsection{NUDD}
\label{subsec:bound-nudd-operator}

If we were to use the same method for NUDD 
{for $m$ qubits} as we did in the QDD case {for one qubit}, we would 
need to consider $2^{(3^m)}$ cases based on the parities of $n_{\vm}$. 
Recall that $n_{\vm}$ indicates how many times 
$\widehat{\sigma}_{\vm}\otimes B_{\vm}$ appears in 
$(\widehat{\sigma}_{\vm_n} \otimes B_{\vm_n})   \cdots   
(\widehat{\sigma}_{\vm_1} \otimes B_{\vm_1})$; Eq.~\eqref{nq_def2}. 
To simplify  the analysis, unlike the QDD case, we 
do not address each error term separately.

Any operator acting on the space of $m$ qubits can be expanded in the 
$\{\sigma_{\vm}\}$ basis.  Generalizing the QDD case [Eq.~\eqref{Usplit}] 
we can thus write
\be
\label{nu-pauli}
U(T)=\sum_{\vm\in {D_m}}\widehat{\sigma}_{\vm}\otimes A_{\vm}(T),
\ee 
where $\widehat{\sigma}_{\vm}$ {and $D_m$ 
were defined in Eqs.~\eqref{sigma_D}}.
Eq.~\eqref{nu-pauli} provides another way to organize the terms in 
Eq.~\eqref{nUseries}. 
We introduce the upper bounding series $S_{\vec{\mu}}$ for each $A_{\vm}$ such that
\bes
\bea
\ma{\|A_{\vm}(T)\|} & \ma{\leq} & \ma{S_{\vec{\mu}}(T)} \label{eq:A<S} \\
S_{\text{N}}(T) &=& \sum_{\vm\in D_m} S_{\vec{\mu}}(T)
\eea
\ees
holds. \ma{Below we construct this series explicitly.} Let
\begin{subequations}
\bea
\vec{0} &:=& \{0,\dots,0\} \\
{\cal K} &:=& {D_m} \backslash \vec{0}.
\eea
\end{subequations}
Since $\widehat{\sigma}_{\vec{0}}$ is the 
all-identity term, it is the term that causes no errors; its
\ma{corresponding} bounding series is $S_{\vec{0}}(T)$. On the other hand, 
every $\widehat{\sigma}_{\vm}$ with $\vm \in {\cal K}$ is an error term \ma{with corresponding bounding series}
$S_{\vm}(T)$. For later use, we note that $S_{\vec{0}}(0)=1$
and $S_{\vm}(0)=0$ for all $\vm \in {\cal K}$.

In order to derive bounds we first recall that $U(T)$ from 
Eq.~\eqref{Udef-nudd} is the solution of the 
Schr\"odinger equation
\bes
\bea
\partial_T U(T) &=&-iH(T) U(T)
\\
\label{U-diff-eq}
 &=&-i\left[ \sum_{\vm\in D_m} f_{\vm}(t/T) \widehat\sigma_{\vm} B_{\vm} \right] U(T)
\eea
\ees
with $U(0)=\mathbb{1}$. Eq.~\eqref{U-diff-eq} generates exactly all
terms of the series of $U(T)$.  Thus, in order to obtain the series
for $S_{\text{N}}(T)$ all we need to do is to replace $-i\to |-i|=1$,
$f_{\vm}\to |f_{\vm}|=1$, 
$\widehat\sigma_{\vm}\to \Vert\widehat\sigma_{\vm}\Vert=1$, and
$B_{\vm}\to\Vert B_{\vm}\Vert=J_{\vm}$ on the 
right hand side. This provides us with the generating differential
equation for the bounding series
\be
\partial_T S_{\text{N}}(T) =
\left[ \sum_{\vm\in D_m}  J_{\vm} \right] S_{\text{N}}(T).
\ee
Its integration from $S_{\text{N}}(0)=1$ recovers the
previous result \eqref{nbseries} precisely.

In analogy, the differential equation for each $A_{\vm}(T)$ reads
\be
\label{A-diff_eq}
\partial_T A_{\vm}(T) = -i\left[ \sum_{\vn\in D_m} f_{\vm\oplus\vn}(t/T) 
R_{\vm\oplus\vn;\vn} B_{\vm\oplus\vn} \right]  A_{\vn}(T),
\ee
where, as before, $\oplus$ stands for sums modulo 2. Note that
addition and subtraction are equivalent modulo 2: $\vm\oplus\vn=
\vm\ominus\vn$. It is essential that the sums modulo 2 in $\{(0,0),(1,0),(1,1),
(0,1)\}$ faithfully reproduce the spin algebra of the identity and the Pauli
matrices except for factors of $\pm i$, which are collected in
$R_{\vm\oplus\vn;\vn}$. Replacing operators by their norms and
complex numbers by their moduli on the right hand side
yields the generating differential equation for the bounding series $S_{\vm}$ \ma{satisfying Eq.~\eqref{eq:A<S},}
\be
\label{S-diff_eq}
\partial_T S_{\vm}(T) = \left[ \sum_{\vn\in D_m}  J_{\vm\oplus\vn} \right] 
S_{\vn}(T).
\ee

Unfortunately, the set of differential equations \eqref{S-diff_eq}
is still too difficult to be solved generally. Hence we aim at
a looser bound by defining
\begin{subequations}
\label{def-J1}
\bea
J_1&:=&\max_{\vm\in \K} J_{\vm}\\
\label{eta2}
\eta&:=&\frac{J_{1}}{J_{\vec{0}}}\\
\label{eps2}
\varepsilon&:=&J_{\vec{0}}T.
\eea
\end{subequations}

If we substitute $J_1$ for each $J_{\vm}$ with $\vm\in\K$ we
obtain one bounding series $S_1(T)$ for all
$ S_{\vm}(T)= S_1(T)$. To see this we insert  $S_1(T)= S_{\vm}(T)$
on the right hand side of Eq.~\eqref{S-diff_eq}
yielding
\bes
\label{bound-diff}
\bea
\label{bound-diff1}
\partial_T S_{\vec{0}}(T) &=&  J_{\vec{0}} S_{\vec{0}}(T) +
\gamma J_1 S_1(T)
\\
\label{bound-diff2}
\partial_T S_{\vm}(T) &=& J_{\vec{0}} S_1(T)+
J_1  S_{\vec{0}}(T) +(\gamma-1)J_1 S_1(T)\qquad
\eea
\ees
where $\vm\in\K$ and $\gamma:=4^m-1=|\K|$ is the number of non-identity terms.
The first term on the right hand side of \eqref{bound-diff1} results from 
$\vec{0}=\vec{0}\oplus\vec{0}$, the second
from $\vn=\vec{0}\oplus \vn$ for $\vn\in\K$.
The first term on the right hand side of \eqref{bound-diff2} results from 
$\vec{0}=\vm\oplus \vm$, the second from 
$\vm=\vm\oplus\vec{0}$, and the third from 
$\vm\oplus\vn$ where $\vn\in\K \backslash \vm$.
There are $\gamma-1$ terms of the latter kind, independent of $\vm$.
This independence implies the equality of all $S_{\vm}(T)$
for $\vm\in\K$ because they all start at the same value $S_{\vm}(0)=0$.
Thus, replacing Eq.~\eqref{bound-diff2}, we have
\be
\label{bound-diff3}
\partial_T S_1(T) = J_{\vec{0}} S_1(T) +
J_1  S_{\vec{0}}(T) +(\gamma-1)J_1 S_1(T)
\ee
which, together with Eq.~\eqref{bound-diff1}, constitutes a two-dimensional 
set of linear differential equations. The two
eigenvalues of the corresponding matrix are $\lambda_+=J_{\vec{0}}+\gamma J_1$
and $\lambda_-=J_{\vec{0}}-J_1$ so that the general solution
reads $S_1(T):=A\exp(\lambda_+T)+B\exp(\lambda_-T)$ with coefficients
$A$ and $B$. The initial
conditions $S_{\vec{0}}(0)=1$ and $S_1(0)=0$ imply $\partial_T S_1(0)=J_1$
from which  $A$ and $B$ are determined to yield
\be
\label{S_1_result}
S_1(T) = 
\frac{\exp(J_{\vec{0}}T)}{\gamma+1}(\exp(\gamma J_1 T)-\exp(-J_1T)).
\ee
The bounding series of the sum of all error terms is given by
$S_{\K}(T):=\gamma S_1(T)$ which reads
\be
\label{S_K_result}
S_{\K}(T) = 
\frac{\gamma\exp(J_{\vec{0}}T)}{\gamma+1}(\exp(\gamma J_1 T)-\exp(-J_1T)).
\ee
This result is the bounding series for all error terms in NUDD.

We define 
\begin{subequations}
\bea
\label{n-taylor}
p_{k}&:=&\frac{1}{k!}\frac{\partial^k}{\partial T^k}S_{\K}{\Big\vert}_{T=0}\\
\label{n-bseries-taylor}
\Delta_N &:=& \sum_{k=N+1}^\infty p_k T^k,
\eea
\end{subequations}
and
\be
d_{\min}:=\min_{1\leq i\leq 2m}\{N_i'\}
\ee
be the decoupling order of the NUDD sequence \cite{Jiang:11}. \green{Since $\Vert A_{\vm}\Vert \leq S_{\vm}(T)$, we have
\be
\sum_{\vm\in \K}\Vert A_{\vm} \Vert \leq \sum_{\vm\in\K} S_{\vm}(T)\leq S_\K(T),
\ee}
\ma{and hence}
\be
\label{nop-bd-res}
\sum_{\vm\in \K}\vv A_{\vm}\vv  \leq  \Delta_{d_{\min}}.
\ee
By Taylor-expanding Eq.~\eqref{S_K_result} we can write $S_{\K}(T)$ and thus 
$\Delta_d$ as an infinite series in terms of $\varepsilon$ and $\eta$ of 
Eqs.~\eqref{eps2} and \eqref{eta2}.
We do not write the result explicitly, but the formula \eqref{S_K_result}
for $S_{\K}(T)$ is simple enough to allow 
the bounds to be computed easily with any computer algebra 
program. This is our first key result for the NUDD sequence.

\begin{figure}[H]
\begin{center}
\includegraphics[scale=.5,width=0.99\columnwidth]
{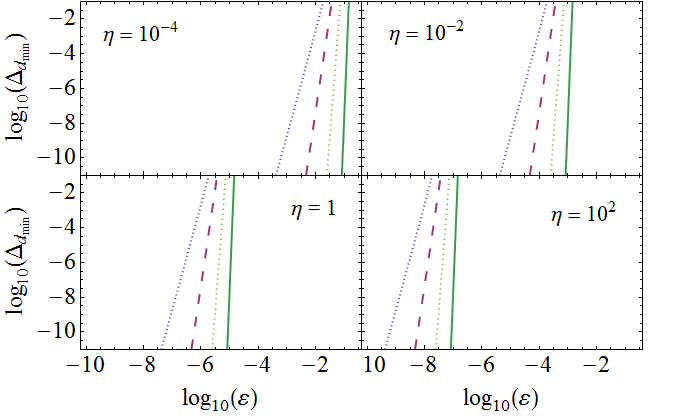} 
\end{center}
\caption{(Color online) The upper bound $\Delta_{d_{\min}}$ for 
$\sum_{\vm\in \K}\vv A_{\vm}\vv$ is given as a function of $\varepsilon$ for 
various pulse numbers $\ma{d_{\min}} \in \{5,10,20,40\}$, 
at fixed values of $\eta$\ma{, and for $m=10$}.  
In each panel the curves become steeper as $d_{\min}$ increases.}
\label{fig:xy3}
\end{figure}

To leading order in $\varepsilon$,  we find
\be
\label{nseries-delta}
\Delta_d = 
g_{d+1}(\eta,m)\varepsilon^{d+1}+\OO(\varepsilon^{d+2}),
\ee
where
\be
\label{ng-def}
g_{l}(\eta,m):=(1-4^{-m})\frac{(1-\eta+4^m\eta)^l-(1-\eta)^l}{l!}.
\ee
In Fig.~\ref{fig:xy3} we include plots of $\Delta_{d_{\min}}$ as a function of 
$\varepsilon$ for 
{various} values of ${d_{\min}}$ and $\eta$. 
 We can see that, as 
expected, the behavior of $\Delta_{d_{\min}}$ is similar to that of $L_\alpha$ 
in the QDD case.

\section{Distance bound}
\label{sec:disbound}
\subsection{QDD}
\label{subsec:D-bound}
Next, we shall use the bounds on the bath operators $A_x,A_y$ and $A_z$ to 
derive a bound on the trace-norm distance 
\be
\label{eq:t-n-dis-def}
D[\rho_{\text{S}}(T),\rho_{\text{S}}^0(T)] = 
\ma{\frac{1}{2} \big\Vert \rho_{\text{S}}(T)-\rho_{\text{S}}^0(T) \big\Vert_1 }
\ee
between the actual qubit state 
\be
\label{qubit}
\rho_{\text{S}}(T):=\tr_{\text{B}}[\rho_{\text{SB}}(T)],
\ee
and the ``error-free'' qubit state
\be
\label{id-qubit}
\rho_{\text{S}}^0(T):=\tr_{\text{B}}[\rho_{\text{SB}}^0(T)]
\ee
where $\rho_{\text{SB}}^0(T)$ is the time-evolved state of the whole qubit-bath 
system  without coupling between the qubit and the bath, 
and $\tr_{\text{B}}$ denotes the 
partial trace over the bath degrees of freedom. \ma{The trace norm $\|A\|_1$ is the trace of $(A^\dagger A)^{1/2}$. The trace-norm distance $D$ is the standard distance measure between density matrices \cite{Nielsen:book}.}
The method we employ here is similar to the one in Ref.~\cite{UL:10}.  
We consider an initial state 
\be
\label{init-state}
\rho_{\text{SB}}^0(0)=\ket{\psi}\bra{\psi}\otimes\rho_{\text{B}}
\ee
in which the qubit is in a pure state $\ket{\psi}$ and the bath is in an 
arbitrary state $\rho_{\text{B}}$ (e.g., a mixed thermal equilibrium state).  The 
initial state evolves to $\rho_{\text{SB}}(T)=U(T)\rho_{\text{SB}}^0(0)U^\dagger(T)$ 
when the qubit and the bath are coupled, or to $\rho_{\text{SB}}^0(T)=
{[\mathbb{1}}_{\text{S}}\otimes U_{\text{B}}(T){]} \rho_{\text{SB}}^0(0)
{[\mathbb{1}}_{\text{S}} \otimes U_{\text{B}}^\dagger(T){]}$ 
when the qubit is isolated from its 
environment.  The unitary bath time-evolution operator without coupling reads
\be
\label{bath-t-evol}
U_{\text{B}}(T):=\exp(-iTB_0)
\ee
where $B_0$ is the bath term in Eq.~\eqref{hamil-gendec1}.

Let us first define the bath correlation functions
\be
\label{bfunc}
b_{\alpha\beta}(T) := \tr
[A_\alpha(T)\rho_{\text{B}} A_{\beta}^\dagger(T)]
\ee
where $\alpha,\beta\in\{0,x,y,z\}$.  To simplify the notation, we will 
omit the time dependence $T$ in $b_{\alpha\beta}(T)$ and $A_\alpha(T)$ from now 
on.  Computation yields (see Appendix~\ref{app:dis-bd-cal})
\bea
\DD &\leq& |b_{xx}+b_{yy}+b_{zz}|\notag \\
& & + \frac{1}{2}
\sum_{\alpha,\beta\in\{0,x,y,z\}}(|b_{\alpha\beta}|-|b_{00}|).
\quad
\label{dis-bd1}
\eea
We know from the unitarity of $U$ and Eq.~\eqref{Usplit} that
\begin{eqnarray}
\nonumber
{\mathbb{1}}&=&U^\dagger(T)U(T)\\
\nonumber
& = & {\mathbb{1}}_{\text{S}}
\otimes\sum_{\alpha\in\{0,x,y,z\}}A_\alpha^\dagger A_\alpha\\
\nonumber
& &\,\,\,\,\,\,+\sum_{\alpha\in\{x,y,z\}}\sigma_\alpha\otimes(A_\alpha^\dagger A_0+A_0^\dagger A_\alpha+i[A_\beta^\dagger,A_\gamma]), \\
\label{unit}
\end{eqnarray}
where in the last sum {$\{\beta,\gamma\}$ 
are adjusted to $\alpha$ so that $\{\alpha,\beta,\gamma\}$ 
is a cyclic permutation of $\{x,y,z\}$.}
Therefore we have 
\begin{subequations}
\begin{eqnarray}
\label{unit-res1}
 {\mathbb{1}} &=& A_0^\dagger A_0+A_x^\dagger A_x+A_y^\dagger A_y+A_z^\dagger A_z,\\
\label{unit-res2}
0 &=& A_x^\dagger A_0+A_0^\dagger A_x+i A_y^\dagger A_z-i A_z^\dagger A_y,\\
\label{unit-res3}
0 &=& A_y^\dagger A_0+A_0^\dagger A_y+i A_z^\dagger A_x-i A_x^\dagger A_z,\\
\label{unit-res4}
0 &=& A_z^\dagger A_0+A_0^\dagger A_z+i A_x^\dagger A_y-i A_y^\dagger A_x.
\end{eqnarray}
\end{subequations}
It follows from Eq.~\eqref{unit-res1} that 
$\bra{i}\sum_{\alpha\in\{0,x,y,z\}}A_\alpha^\dagger A_\alpha\ket{i} = 1$
for all normalized states $\ket{i}$, and thus
\be
\label{unit-inq1}
\bra{i}A_0^\dagger A_0\ket{i} = \Vert A_0\ket{i}\Vert^2 \leq 1
\ee
because $\bra{i}\sum_{\alpha\in\{x,y,z\}}A_\alpha^\dagger A_\alpha\ket{i} = 
\sum_{\alpha\in\{x,y,z\}} \Vert A_\alpha\ket{i}\Vert^2$ is nonnegative.  
In particular, we know that 
\be
\label{unit-inq2}
\Vert A_0 \Vert \leq 1.
\ee

To obtain a bound on the functions $b_{\alpha\beta}$ we use the following 
general correlation functions inequality (see Ref.~\cite{UL:10} 
for a proof)
\be
\label{cor-func-inq}
|\tr[Q\rho_{\text{B}} Q']|\leq\Vert Q\Vert \Vert Q'\Vert,
\ee
which holds for arbitrary bounded bath operators $Q,Q'$.  
Using Eq.~\eqref{unit-inq2} , Eq.~\eqref{cor-func-inq}, and the bounds 
\eqref{eq:Aperp} in Eq.~\eqref{dis-bd1} yields 
\begin{subequations}
\begin{eqnarray}
\nonumber
\lefteqn{\DD}\\
\nonumber
&\leq& \vv A_x\vv+\vv A_y\vv+\vv A_z\vv+\Vert A_x\Vert^2+
\Vert A_y\Vert^2+\Vert A_z\Vert^2\\
\label{dis-bd-res1}
& &{}+\vv A_x\vv\vv A_y\vv+\vv A_y\vv\vv A_z\vv+\vv A_x\vv\vv A_z\vv\\
\nonumber
&\leq& L_x+L_y+L_z+L_x^2+L_y^2+L_z^2+L_xL_y+L_yL_z\\
\label{dis-bd4}
& &{}+L_xL_z\\
\label{dis-bd-res2}
&=&\Delta_{d_x}^{(3)} + \Delta_{d_x}^{(4)}+\Delta_{d_y}^{(2)}+ \Delta_{d_y}^{(5)}+
\Delta_{d_z}^{(1)} + \Delta_{d_z}^{(6)}\\
\nonumber
& &{}+(\Delta_{d_x}^{(3)})^2+ (\Delta_{d_x}^{(4)})^2+(\Delta_{d_y}^{(2)})^2+ 
(\Delta_{d_y}^{(5)})^2+(\Delta_{d_z}^{(1)})^2\\
\nonumber
& &{} + (\Delta_{d_z}^{(6)})^2+2\Delta_{d_x}^{(3)}\Delta_{d_x}^{(4)}+
2\Delta_{d_y}^{(2)}\Delta_{d_y}^{(5)}+2\Delta_{d_z}^{(1)} \Delta_{d_z}^{(6)}\\
\nonumber
& &{}+\Delta_{d_x}^{(3)}\Delta_{d_y}^{(2)}+\Delta_{d_x}^{(3)}\Delta_{d_y}^{(5)}+
\Delta_{d_x}^{(4)}\Delta_{d_y}^{(2)}+ \Delta_{d_x}^{(4)}\Delta_{d_y}^{(5)}\\
\nonumber
& &{}+\Delta_{d_y}^{(2)}\Delta_{d_z}^{(1)}+\Delta_{d_y}^{(2)}\Delta_{d_z}^{(6)}+
\Delta_{d_y}^{(5)}\Delta_{d_z}^{(1)}+ \Delta_{d_y}^{(5)}\Delta_{d_z}^{(6)}\\
\nonumber
& &{}+\Delta_{d_x}^{(3)}\Delta_{d_z}^{(1)}+\Delta_{d_x}^{(3)}\Delta_{d_z}^{(6)}+
\Delta_{d_x}^{(4)}\Delta_{d_z}^{(1)}+ \Delta_{d_x}^{(4)}\Delta_{d_z}^{(6)}.
\end{eqnarray}
\end{subequations}
This is the rigorous bound on the trace-norm distance and hence
our second key result for QDD.

Using Eq.~\eqref{delta-poly2}, one can rewrite this
bound in terms of $\varepsilon$ and $\vec{\eta}$,
which we do not write out here for brevity.

In explicit calculations the leading order in $\varepsilon$ will dominate
for $\varepsilon \to 0$.
Then only the first line in Eq.~\eqref{dis-bd-res2} plays a 
role, since all other terms are of higher order. Hence we have
\begin{eqnarray}
\nonumber
\DD &\le&
[g_{d_x+1}^{(3)}(\vec{\eta})+g_{d_x+1}^{(4)}(\vec{\eta})]\varepsilon^{d_x+1} 
\nonumber\\
&&  + [g_{d_y+1}^{(2)}(\vec{\eta})+g_{d_y+1}^{(5)}(\vec{\eta})]
\varepsilon^{d_y+1} \nonumber\\
&& + [g_{d_z+1}^{(1)}(\vec{\eta})+g_{d_z+1}^{(6)}(\vec{\eta})] 
\varepsilon^{d_z+1} \nonumber \\
\label{rig-dis-bd}
&& +\OO(\varepsilon^{\min_{\alpha\in\{x,y,z\}}\{d_\alpha\}+2}).
\end{eqnarray}
Equation~\eqref{rig-dis-bd} is
our third key QDD result. It shows that the QDD bound 
is dominated by the channel with the smallest decoupling order $d_\alpha$,
as was expected intuitively.

\subsection{NUDD}
\label{subsec:nD-bd}

We consider an initial state 
\be
\label{ninit-state}
\rho_{\text{SB}}^0(0) = \ket{\psi}\bra{\psi} \otimes \rho_{\text{B}}
\ee
where $\ket{\psi}$ is the state of the $m$-qubit system.  The initial state 
evolves to $\rho_{\text{SB}}(T)=U(T)\rho_{\text{SB}}^0(0)U^\dagger(T)$ 
when the qubit and the bath are coupled, or to $\rho_{\text{SB}}^0(T)=
{[\mathbb{1}_{\text{S}}\otimes U_{\text{B}}(T)]\rho_{\text{SB}}^0(0)
[\mathbb{1}_{\text{S}} \otimes U_{\text{B}}^\dagger(T)]}$ 
when the qubit is isolated from its environment.  
The unitary bath time-evolution operator without coupling reads
\be
\label{nbath-t-evol}
U_{\text{B}}(T):=\exp(-iTB_{\vec{0}})
\ee
where $B_{\vec{0}}$ is the bath operator that corresponds to the identity system 
operator.  
We know from the unitarity of $U(T)$ and Eq.~\eqref{nu-pauli} that 
\be
\label{nunitarity}
\sum_{\vm\in {D_m}}A_{\vm}^\dagger A_{\vm}={\mathbb{1}}.
\ee
Following the same steps as in the QDD case yields
\be
\label{n-bd-A}
\vv A_{\vec{0}}\vv\leq 1.
\ee
Explicit computation (see Appendix~\ref{app:ndis-bd-cal}) as in the QDD case 
shows that
\be
\DD 
\label{ndis-bd-operator}
\leq \Big(\sum_{\vm\in {\K}} \Vert A_{\vm}\Vert\Big)^2+
\Vert A_{\vec{0}}\Vert\sum_{\vm\in {\K}} \Vert A_{\vm}\Vert.
\ee
We use the bounds on the operators $A_{\vm}$ to derive a bound on the 
trace norm distance between $\rho_{\text{SB}}(T)$ and $\rho_{\text{SB}}^0(T)$.  
Substituting Eq.~\eqref{nop-bd-res} and Eq.~\eqref{n-bd-A} into 
Eq.~\eqref{ndis-bd-operator}, we conclude that 
\be
\label{ndis-bd}
\DD\leq \Delta_{d_{\min}}^2+\Delta_{d_{\min}},
\ee
where $\Delta_{d_{\min}}$ is the bound presented in Eq.~\eqref{n-bseries-taylor}.
The bound is dominated by $\Delta_{d_{\min}}$
 which is the leading order term for $\varepsilon\to 0$.  By expanding
$\Delta_{d_{\min}}$ we find for the leading order term
\be
\label{ndis-bd-rig}
\DD \le g_{{d_{\min}}+1}(\eta,m)\varepsilon^{{d_{\min}}+1}+
{\OO(\varepsilon^{{d_{\min}}+2})}.
\ee
Equation~\eqref{ndis-bd-rig} is our second key result for NUDD.

\section{Conclusions}
\label{sec:conclusions}

We have derived and presented rigorous performance bounds for the QDD and 
NUDD sequences, respectively. These sequences, which build on the UDD sequence, 
protect a qubit or system of qubits from general decoherence, under the 
assumptions that the pulses are instantaneous and the bath operators are 
bounded in operator norm. Our key bounds are given 
in Eq.\ \eqref{dis-bd-res2}
for QDD and in  Eq.~\eqref{ndis-bd} for NUDD. 
The leading order terms are identified  in Eq.~\eqref{rig-dis-bd} 
(for QDD) and Eq.~\eqref{ndis-bd-rig} (for NUDD).  These 
results show that if  the total sequence time is fixed, we can make the error 
$\DD$ arbitrarily  small by increasing the number of pulses at each 
UDD level comprising the  QDD or NUDD sequences. 

When instead the minimum pulse interval is fixed, we expect that, just as 
in the case of UDD \cite{UL:10}, there will be an optimal sequence order, 
beyond which performance starts to decrease. A rigorous study of this aspect of
 QDD and NUDD is an interesting topic for a future investigation. We hope that 
the results presented here will inspire experimental tests of the QDD and NUDD 
sequences in physical systems with bath spectral densities exhibiting 
relatively hard high frequency cutoffs, a condition which corresponds to our 
key assumptions [Eqs.~\eqref{Jbounds},\eqref{nbound}] of 
bounded bath operators.

\begin{acknowledgments}

We are grateful to Wan-Jung Kuo and Gerardo A. Paz-Silva for helpful 
discussions and comments.  Y.X.\ thanks the USC Center for Quantum Information 
Science \& Technology, where this work was done, and the Science Horizons 
Research Fellowships from Howard Hughes Medical Institute and 
Bryn Mawr College for financial support.  G.S.U.\  is supported under 
DFG grant UH 90/5-1. D.A.L.\ and Y.X.\ are both supported by the NSF 
Center for Quantum Information and Computation
for Chemistry, award number CHE-1037992. D.A.L.\ is also sponsored by the 
United States Department of Defense.
The views and conclusions contained in this document are those of the 
authors and should not be interpreted as representing the official policies, 
either expressly or implied, of the U.S.\ Government. This research is partially supported by the ARO MURI grant
W911NF-11-1-0268.
\end{acknowledgments}

\appendix

\section{Bounding polynomials}
\label{app:g}
The polynomials that appear in the Taylor expansions of the 
bounding functions $S_j(\varepsilon,\vec{\eta})$ are:
\begin{eqnarray}
\nonumber
g_l^{(1)}(\vec{\eta})  &=& \frac{1}{8l!} [(1+\as+\bb+\cc)^l+(1-\as+\bb+\cc)^l\\
\nonumber
& &{}+(1+\as-\bb+\cc)^l+(1-\as-\bb+\cc)^l\\
\nonumber
& &{}-(1+\as+\bb-\cc)^l-(1-\as+\bb-\cc)^l\\
\nonumber
& &{}-(1+\as-\bb-\cc)^l-(1-\as-\bb-\cc)^l]\\
\label{bseries1}
\eea
\bea
\nonumber
g_l^{(2)}(\vec{\eta})& = & \frac{1}{8l!}[(1+\as+\bb+\cc)^l+(1-\as+\bb+\cc)^l\\
\nonumber
& &{}-(1+\as-\bb+\cc)^l-(1-\as-\bb+\cc)^l\\
\nonumber
& &{}+(1+\as+\bb-\cc)^l+(1-\as+\bb-\cc)^l\\
\nonumber
& &{}-(1+\as-\bb-\cc)^l-(1-\as-\bb-\cc)^l]\\
\label{bseries2}
\eea
\bea
\nonumber
g_l^{(3)}(\vec{\eta}) & = & \frac{1}{8l!}[(1+\as+\bb+\cc)^l+(1-\as+\bb+\cc)^l\\
\nonumber
& &{} -(1+\as-\bb+\cc)^l-(1-\as-\bb+\cc)^l\\
\nonumber
& & {}-(1+\as+\bb-\cc)^l-(1-\as+\bb-\cc)^l\\
\nonumber
& & {}+(1+\as-\bb-\cc)^l+(1-\as-\bb-\cc)^l]\\
\label{bseries3}
\eea
\bea
\nonumber
g_l^{(4)}(\vec{\eta}) & = & \frac{1}{8l!}[(1+\as+\bb+\cc)^l -(1-\as+\bb+\cc)^l\\
\nonumber
& &{} + (1+\as-\bb+\cc)^l - (1-\as-\bb+\cc)^l\\
\nonumber
& &{} + (1+\as+\bb-\cc)^l - (1-\as+\bb-\cc)^l\\
\nonumber
& &{} + (1+\as-\bb-\cc)^l - (1-\as-\bb-\cc)^l]\\
\label{bseries4}
\eea
\bea
\nonumber
g_l^{(5)}(\vec{\eta}) & = & \frac{1}{8l!}[(1+\as+\bb+\cc)^l-(1-\as+\bb+\cc)^l\\
\nonumber
& &{} +(1+\as-\bb+\cc)^l-(1-\as-\bb+\cc)^l \\
\nonumber
& &{} -(1+\as+\bb-\cc)^l+(1-\as+\bb-\cc)^l\\
\nonumber
& &{} -(1+\as-\bb-\cc)^l+(1-\as-\bb-\cc)^l]\\
\label{bseries5}
\eea
\bea
\nonumber
g_l^{(6)}(\vec{\eta}) & = &\frac{1}{8l!}[(1+\as+\bb+\cc)^l-(1-\as+\bb+\cc)^l\\
\nonumber
& &{} -(1+\as-\bb+\cc)^l+(1-\as-\bb+\cc)^l \\
\nonumber
& &{} +(1+\as+\bb-\cc)^l-(1-\as+\bb-\cc)^l\\
\nonumber
& &{} -(1+\as-\bb-\cc)^l+(1-\as-\bb-\cc)^l]\\
\label{bseries6}
\end{eqnarray}

\section{Distance Bound Calculation}
\subsection{QDD}
\label{app:dis-bd-cal}
In this appendix we prove the bound on the trace-norm distance in 
Eq.~\eqref{dis-bd1}. To simplify the notation, we let 
\be
P_0:=\is .
\ee
Then
\begin{subequations}
\begin{eqnarray}
\nonumber
\lefteqn{2\DD} \\
\nonumber
&=&\big\Vert\tr_{\text{B}}\left[\rho_{\text{SB}}(T)-\rho_{\text{SB}}^0(T)
\right]\big\Vert_1\\
\nonumber
&=&\big\Vert\tr_{\text{B}}\left[U(T)\rho_{\text{SB}}^0(0)U^\dagger(T)\right]-
\tr_{\text{B}}\left[\rho_{\text{SB}}^0(0)\right]\big\Vert_1\\
\nonumber
&=&\left\|\tr_{\text{B}}\left\{\left[\sum_{\alpha\in\{0,x,y,z\}}\sigma_\alpha\otimes 
A_\alpha\right](P_0\otimes\rho_{\text{B}}) \right.\right.\\
\nonumber
& &\left.\left.\left[\sum_{\alpha\in\{0,x,y,z\}}\sigma_\alpha\otimes 
A_\alpha^\dagger\right]\right\}-P_0\right\|_1\\
\nonumber
&=&\big\Vert (b_{00}-1)P_0+b_{0x}P_0\sigma_x+b_{0y}P_0\sigma_y\\
\nonumber
& &{}+b_{0z}P_0\sigma_z+b_{x0}\sigma_xP_0+b_{xx}\sigma_xP_0\sigma_x\\
\nonumber
& &{}+b_{xy}\sigma_xP_0\sigma_y+b_{xz}\sigma_xP_0\sigma_z+b_{y0}\sigma_yP_0\\
\nonumber
& &{}+b_{yx}\sigma_yP_0\sigma_x+b_{yy}\sigma_yP_0\sigma_y+b_{yz}\sigma_yP_0\sigma_z\\
\nonumber
& &{}+b_{z0}\sigma_zP_0+b_{zx}\sigma_zP_0\sigma_x+b_{zy}\sigma_zP_0\sigma_y\\
\label{tracenorm1}
& &{}+b_{zz}\sigma_zP_0\sigma_z\big\Vert_1\\
\nonumber
&\leq&|b_{00}-1|+|b_{0x}|+|b_{0y}|+|b_{0z}|\\
\nonumber
& &{}+|b_{x0}|+|b_{xx}|+|b_{xy}|+|b_{xz}|\\
\nonumber
& &{}+|b_{y0}|+|b_{yx}|+|b_{yy}|+|b_{yz}|\\
\label{tracenorm2}
& &{}+|b_{z0}|+|b_{zx}|+|b_{zy}|+|b_{zz}|.
\end{eqnarray}
\end{subequations}
In going from Eq.~\eqref{tracenorm1} to Inq.~\eqref{tracenorm2}, 
we used the triangle inequality, the unitary invariance of the trace norm, 
and the normalization of $\ket{\psi}$. This proves Eq.~\eqref{dis-bd1}.

We require one last result:  
\be
\label{b-identity}
\begin{split}
&|b_{00}-1|= |\tr[A_0\rho_{\text{B}} A_0^\dagger]-1|\\
&=|\tr[A_0^\dagger A_0\rho_{\text{B}}]-1|\\
&=|\tr\{[\mathbb{1}-A_x^\dagger A_x-A_y^\dagger A_y
-A_z^\dagger A_z]\rho_{\text{B}}\}-1|\\
&=|\tr[\rho_{\text{B}}]-\tr[A_x\rho_{\text{B}} A_x^\dagger]
-\tr[A_y\rho_{\text{B}} A_y^\dagger]-
\tr[A_z \rho_{\text{B}} A_z^\dagger]|\\
&=|b_{xx}+b_{yy}+b_{zz}|,
\end{split}
\ee
where we used cyclic invariance of the trace in 
$b_{\alpha\beta}$ together with Eq.~\eqref{unit-res1} and the normalization 
$\tr[\rho_{\text{B}}]=1$. 
Eq.~\eqref{dis-bd1} now follows immediately from the triangle inequality.

\subsection{NUDD}
\label{app:ndis-bd-cal}
Here we prove Eq.~\eqref{ndis-bd-operator} following the same steps as in 
the QDD case.
\bes
\begin{eqnarray}
\nonumber
\lefteqn{\DD}\\
\nonumber
&=&\big\Vert\tr_{\text{B}}[\rho_{\text{SB}}(T)-\rho_{\text{SB}}^0(T)]\big\Vert_1\\
\nonumber
&=&\big\Vert\tr_{\text{B}}[U(T)\rho_{\text{SB}}^0(0)U^\dagger(T)]-
\tr_{\text{B}}[\rho_{\text{SB}}^0(0)]
\big\Vert_1\\
\nonumber
&=&\left\|\tr_{\text{B}}\left\{\left[\sum_{\vm\in\{0,1\}^{2m}}\sigma_{\vm}A_{\vm}\right]
(P_0 \otimes\rho_{\text{B}})\right.\right. \\ \notag 
&&\left.\left.\left[\sum_{\vm\in\{0,1\}^{2m}}\sigma_{\vm}A_{\vm}^\dagger\right]\right\}-
P_0\right\|_1\\
\nonumber
&=&\left\|\sum_{\vm\in\{0,1\}^{2m}} \sum_{\vn\in\{0,1\}^{2m}} 
\tr(A_{\vm}^\dagger\rho_{\text{B}} A_{\vn})\sigma_{\vm} P_0 \sigma_{\vn}-P_0\right\|_1\\
\nonumber
&\leq&|\tr(A_{\vec{0}}^\dagger \rho_{\text{B}} 
A_{\vec{0}})-1|+\sum_{\vm\in {\cal K}} 
\sum_{\vn\in {\cal K}} |\tr(A_{\vm}^\dagger\rho_{\text{B}} A_{\vn})|\\
\nonumber
& &{}+\sum_{\vm\in {\cal K}} {|}\tr(A_{\vec{0}}^\dagger \rho_{\text{B}} A_{\vm}){|} + 
\sum_{\vm\in {\cal K}} {|}\tr(A_{\vm}^\dagger \rho_{\text{B}} A_{\vec{0}}){|}\\
\nonumber
&\leq& \sum_{\vm\in {\cal K}}|\tr(A_{\vm}^\dagger \rho_{\text{B}} A_{\vm})|+\sum_{\vm\in {\cal K}} 
\sum_{\vn\in {\cal K}} |\tr(A_{\vm}^\dagger\rho_{\text{B}} A_{\vn})|\\
\nonumber
& &{}+\sum_{\vm\in {\cal K}} |\tr(A_{\vec{0}}^\dagger \rho_{\text{B}} A_{\vm})| + 
\sum_{\vm\in {\cal K}} |\tr(A_{\vm}^\dagger \rho_{\text{B}} A_{\vec{0}})|\\
\label{n-op1}
&{\leq}&  {\Big(\sum_{\vm\in \K} \Vert A_{\vm}\Vert\Big)^2
+\sum_{\vm\in \K} \Vert A_{\vm}\Vert^2
+2 \Vert A_{\vec{0}}\Vert\sum_{\vm\in \K} \Vert A_{\vm}\Vert
\qquad
}
\\
\label{n-op}
&\leq& 2 \Big(\sum_{\vm\in {\cal K}} \Vert A_{\vm}\Vert\Big)^2+2 
\Vert A_{\vec{0}}\Vert\sum_{\vm\in {\cal K}} \Vert A_{\vm}\Vert.
\end{eqnarray}
\ees
In the derivation we used the unitary invariance of the trace norm, triangle 
inequality, the normalization of $\ket{\psi}$, Eq.~\eqref{cor-func-inq} and 
Eq.~\eqref{nunitarity}. 
The step from inequality
\eqref{n-op1} to \eqref{n-op} is not necessary, but loosens the bound
for the sake of simplicity of the result.


%

\end{document}